\documentstyle[eqsecnum,aps]{revtex}
\begin{document}
\draft
\title{A Chiral Effective Field Theory and Radiative Decays of Mesons}
\author{ E.Gedalin\thanks{gedal@bgumail.bgu.ac.il},
 A.Moalem\thanks{moalem@bgumail.bgu.ac.il}
 and L.Razdolskaya\thanks{ljuba@bgumail.bgu.ac.il}}
\address
{ Department of Physics, Ben Gurion University, 84105,
Beer Sheva, Israel}
\maketitle
\begin{abstract}An extended  $U(3)_L\bigotimes U(3)_R$ chiral effective
field theory which includes pseudoscalar and vector meson nonets as dynamic
variables is presented. The theory combines a hidden local symmetry
approach with a general procedure of including the $\eta'$ meson into
chiral theory, and accounts
for $direct$ and $indirect$ symmetry breaking effects 
via a mechanism based on the quark mass matrix. 
The theory is applied to anomalous radiative decays of pseudoscalar
and vector mesons, using particle mixing
schemes corresponding to different symmetry breaking assumptions
and uniquely determined by the lagrangian presumed.
Radiative decays of light
flavor mesons are best explained within the framework of a one  mixing
angle scheme and provide evidence for $SU(3)_F$ and nonet symmetry
breaking.
\\
 \end{abstract}
\bigskip

\section{Introduction}
The decays of light flavor mesons have been discussed vigorously in the
literature \cite{gilman87}-\cite{gedalin00}. 
Particularly, phenomenological models based on effective
field theories have been rather successful to explain anomalous and
non-anomalous processes involving vector and pseudoscalar mesons, tensor
and higher-spin mesons, $J/\psi$ decays into a vector and a pseudoscalar
meson, and many other related decays 
and topics\cite{bijnens90,bramon95,feld98b,benayoun981,benayoun99}. 
Particularly successful are effective field theories based on the 
Hidden Local Symmetry (HLS) approach\cite{bando}, claiming to provide 
a consistent framework  which explains a vast amount of
data, and more importantly, reflects on the energy properties of the
more fundamental  QCD lagrangian.

In the chiral limit, the QCD lagrangian exhibits an $SU(3)_L\bigotimes
SU(3)_R$ symmetry which breaks down spontaneously to  $SU(3)_V$,
giving rise to an octet of pseudoscalar light Goldstone
bosons. The QCD spectrum  contains a ninth singlet boson because the 
axial $U(1)$ symmetry is broken by the anomaly. Nowadays it is well 
accepted that the lightest pseudoscalar mesons 
($\pi^+ , \pi^- ,~\pi^0,~\mbox{K}^+, ~\mbox{K}^- , ~\mbox{K}^0 , ~
\bar{\mbox{K}}^0~\mbox{and} ~\eta$) are candidates of the Goldstone  octet.
Though considerably heavier, the $\eta ' (957~MeV)$ meson is considered
to be the pseudoscalar singlet.

Our main interest in the present work is to develop and apply
a $U(3)_L\bigotimes U(3)_R$ chiral theory where the
the pseudoscalar and vector meson nonets play the role of dynamic 
degrees of freedom. The theory combines the HLS
approach of Bando et  al.\cite{bando} with a general 
procedure of  including the $\eta'$ meson into a chiral
theory\cite{gasser85,leut96,leut97,herera97}, and is constructed in close analogy
with QCD. The $SU(3)_L\bigotimes SU(3)_R$ local symmetry based QCD lagrangian is
extended by adding a term proportional to the topological charge operator
(the so called winding number density), which gives rise to the
well known Wess-Zumino-Witten (WZW) anomaly term and,
renders the lagrangian to be $U(3)_L\bigotimes U(3)_R$ locally
symmetric\cite{gasser85}.
Most importantly, the effective lagrangian constructed, exhibits the
fundamental
symmetries of this extended QCD lagrangian and accounts
for  $direct$ and $indirect$ symmetry breaking effects
via a mechanism based on the quark mass matrix.
A natural way to account for $SU(3)$ flavor symmetry breaking terms
involves an expansion of the QCD generating functional in powers of the
quark masses. Here as well we use a similar expansion to generate symmetry
breaking terms. We believe that the
theory proposed defines an accurate framework for electroweak and strong
interactions of light flavor mesons and can be extended easily to include
tensor and higher spin mesons as well as baryons. Although, the structure of the
lagrangian seems similar to the original model of Bando, Kugo,
Yamawaki\cite{bando} and variants of this
model\cite{bramon95,benayoun981,benayoun98},
it includes additional terms due to including the $\eta '$ and the $\phi$ mesons as
dynamical variables. Namely, the trace of the nonet field matrices are
nonvanishing and therefore, the symmetric as well as the symmetry breaking 
lagrangian parts must include the square of the traces of vector and
axial-vector covariants, in addition to terms proportional to traces of
covariants square. These additional terms play an important role in
determining the particle mixing schemes and the meson mass spectra.
In marked difference  with previous 
studies\cite{bramon95,benayoun981,benayoun98}
the meson masses, mixing angles and other observables like radiative decay
widths or form factors are calculated straightforwardly from the
lagrangian presumed without any additional assumptions. The theory guarantees
reproducing the meson masses and as a result the mixing angles are 
functions of the symmetry breaking scales and not free model parameters.

Our paper is organized as follows. In section II we 
construct the lagrangian. In order to  be fully consistent with QCD,
a symmetry breaking companion is added to each of the unbroken
lagrangian terms. These symmetry breaking companions
affect the kinetic and mass terms of the  lagrangian which, in turn
determine the mixing schemes of state and coupling constants 
\cite{gedalin01}. Two alternative ways
are proposed, corresponding to  
$SU(3)_F$ symmetry breaking ($Alternative ~I$) and $U(3)_F$ symmetry
breaking ($Alternative ~II$). The strong condition of nonet symmetry 
often used to relate the octet and singlet pseudoscalar 
states\cite{bramon90,ball96,bramon99,veno98,feld98} 
is not presumed $a~priori$.
In the limit of nonet symmetry (in the sense
of equal singlet and octet radiative decay constants, $F_0 = F_8 $)
the $Alternative ~~II$ scheme reduces to a scheme equivalent to that 
of the quark flavor basis (QFB) of Feldmann
et al. \cite{feld98,feld98a,feld98b}. 
The realization of these mixing schemes is described in section III.
In section IV we apply our model to study anomalous processes.
Namely, we calculate radiative  decay widths of  $V^0 \to P^0 \gamma$,
$P^0 \to V^0 \gamma$ and $P^0 \to \gamma \gamma$, 
with $P^0 = \pi , K, \eta ,\eta '$ and $V^0 =  
\rho , K^*, \omega ,\phi$. Global fit to data is performed in order to
determine numerically the symmetry breaking scales and 
evaluate the success of the different schemes to explain data.
We summarize and conclude in section V.

\section{The effective lagrangian}

The main objective in the present section is to construct 
a most general $U(3)_L\bigotimes U(3)_R$ local symmetry based
effective lagrangian, 
for pseudoscalar and vector meson nonets interacting with external
electroweak fields. As a non-linear representation of the pseudoscalar 
nonet fields we define \cite{gasser85,leut96},
\begin{equation} 
        U(P,\eta_0+F_0\vartheta) \equiv \xi^2(P,\eta_0+F_0\vartheta)
        \equiv \exp \left\{i\frac{\sqrt{2}}{F_8}P +
        i\sqrt{\frac{2}{3}}\frac{1}{F_0}\eta_0 {\bf1}\right\}~,
        \label{pmfield}
\end{equation}
where $\eta_0(x)$ stands for the pseudoscalar singlet and $P(x)$ for the
pseudoscalar Goldstone octet matrix,
 \begin{equation}
   P = \left(\begin{array}{ccc}
	\frac{1}{\sqrt{2}}\pi^0+\frac{1}{\sqrt{6}}\eta_8 & \pi^+ &K^+
\\
	\pi^- & - \frac{1}{\sqrt{2}}\pi^0+\frac{1}{\sqrt{6}}\eta_8  
	&K^0  \\
	K^- & \bar{K}^0 &-\frac{2}{\sqrt{6}}\eta_8
	\end{array}\right)~,
	\label{poctet}
	\end{equation}
with obvious notation.
In this representation, the octet degrees of freedom are contained 
in the unimodular part of the field $U$ while the phase
$detU = \exp i (\sqrt{6}\eta_0/F_0 )$ involves the singlet only.
The vacuum angle serves here as an auxiliary field $\vartheta (x)$
which renders the variable $X(x) \equiv -i\ln detU +\vartheta 
\equiv \sqrt{6}\eta_0(x)/F_0 +\vartheta(x)$ to be invariant under 
$U(3)_L\bigotimes U(3)_R$ transformations \cite{gasser85,leut96}.
Since there exists no dimension-nine 
representation for $U(3)_L\bigotimes U(3)_R$, 
the octet ($F_8$) and singlet ($F_0$)
radiative decay constants may well be different. We follow  Refs.
\cite{gasser88,krause90,bernard95} and define vector 
and axial-vector covariants,
\begin{eqnarray}
	&& \Gamma_\mu =  
   \frac{i}{2}\left[\xi^\dagger ,\partial_\mu\xi\right]
	  +\frac{1}{2}\left(\xi^\dagger r_\mu \xi + 
           \xi l_\mu\xi^\dagger\right) ~,
	\label{gammac} \\
	&&\Delta_{\mu} =
     \frac{i}{2}\left\{\xi^\dagger ,\partial_{\mu}\xi\right\}
  +\frac{1}{2}\left(\xi^\dagger r_\mu\xi - \xi l_\mu\xi^\dagger\right) ~.
	\label{deltaf} 
    \end{eqnarray}
Here $r_\mu$ and $l_\mu$ represent the standard model external gauge
fields; $r_\mu= v_\mu + a_\mu$ and $ \quad l_\mu = v_\mu -
a_\mu$, with $v_\mu$ and  $a_\mu$ being the vector and axial 
vector external electroweak fields, respectively. 
For pure electromagnetic interactions ~~ $l_\mu = r_\mu = 
-eQA_\mu$ where $A_\mu$ denotes the electromagnetic field and 
$Q=diag(2/3, -1/3, -1/3)$ is the quark charge operator.

Under $U(3)_L\bigotimes U(3)_R$ the field, Eqn. \ref{pmfield}, 
transforms as,
  \begin{equation}
  	U' = RUL^\dagger, 
  \end{equation}
with $~R\in U(3)_R, \quad L\in U(3)_L$. The vector ($\Gamma _\mu$) and
axial-vector ($\Delta_\mu$) covariants transform, respectively, 
as gauge and  matter fields, i.e.,
  \begin{eqnarray}
  	 & & \Gamma '_\mu = K \Gamma_\mu K^\dagger + iK\partial_\mu 
  K^\dagger ~,
  	 \\
  	 & & \Delta' _\mu = K \Delta_\mu K^\dagger~,
  \end{eqnarray}
  where $K(U,R,L)$ is a compensatory field representing an element of 
  the conserved vector subgroup
$U(3)_V$\cite{gasser88,krause90,bernard95}.   
  
The dynamical gauge bosons are defined as  a 
  $3 \times 3$ vector field matrix $V_\mu$ which transforms as, 
  \begin{equation}
 V'_\mu =  KV_\mu K^\dagger + \frac{i}{g} K\partial_\mu K^\dagger \ .  
  \end{equation}
Clearly, the vector $\Gamma_\mu -g V_\mu$ as well as the axial vector 
$\Delta_\mu$ transform homogeneously, and to lowest order (i.e. with 
the smallest number of derivatives) the lagrangian  can be constructed 
from the traces of $\Delta^2_\mu$ ,~~ $(\Gamma_\mu - gV_\mu)^2$, ~~
$\Delta_\mu$,~~ $(\Gamma_\mu - gV_\mu)$,~~ $D^\mu \vartheta$, 
and arbitrary functions of the variable $X(x)$ all of which 
being invariant under $U(3)_L\bigotimes U(3)_R$ transformations.
Then to lowest order, a most general form of a symmetric effective
chiral lagrangian is,
\begin{equation}
	L = L_A + aL_V - \frac{1}{4}Tr(V_{\mu\nu}V^{\mu\nu})~ ,
	\label{lag}
\end{equation}
with,
\begin{eqnarray}
     && L_{A}= W_{1}(X)Tr(\Delta_\mu\Delta^\mu) 
	+W_{4}(X)Tr(\Delta_\mu)Tr(\Delta^\mu) +
	 \nonumber\\
	 && W_{5}(X)Tr(\Delta_\mu)D^\mu \vartheta +
	 W_{6}(X)D_\mu \vartheta D^\mu \vartheta ~,
	\label{lagina}\\
	 &  & L_V =
	 \tilde{W}_{1}(X)Tr([\Gamma_\mu - gV_\mu][\Gamma^\mu - gV^\mu])+
	 \nonumber\\ 
	 &  & \tilde{W}_{4}(X)Tr(\Gamma_\mu - gV_\mu)Tr(\Gamma^\mu -
     gV^\mu)~,
	 \label{laginv}
\end{eqnarray}
and,
\begin{eqnarray}
	 &  & 	D_\mu \vartheta = \partial_\mu \vartheta + Tr(r_\mu - l_\mu)~,
	\\
	 &  & V_{\mu\nu} = \partial_\mu V_\nu - \partial_\nu V_\mu -ig 
	 [V_\mu,V_\nu] ~.
\label{vmunu}
\end{eqnarray}
All three terms of the lagrangian $L$ in Eqns.\ref{lag}-\ref{laginv} 
are invariant under $U(3)_L\bigotimes U(3)_R$ transformations.

The coefficient functions, $W_i, ~\mbox{and}~\tilde{W}_i$, are constrained
by parity conservation to be even functions of the
variable $X$. In addition by requiring that the
normalization of the HLS kinetic term be equal 1/2, 
it is easy to show that  $W_{1}(0) = F^2_8, \quad 
W_{4}(0) = (F^2_0 - F^2_8)/3,~~\mbox{and}  \quad W_{6}(0) = 1/2$. 
Such a normalization ensures that the pseudoscalar 
singlet couples to the singlet axial current with a strength $F_0$ while
the octet states couple to the octet axial currents with a strength 
$F_8$. 

Although the lagrangian, Eqn.\ref{lag}, appears similar to that of 
Bando et al. \cite{bando}, the terms $L_A$ and $L_V$ are different. Namely, 
including the $\eta '$ and $\phi$ mesons as  dynamical variables involves
additional terms with 
$Tr (\Delta_{\mu})Tr (\Delta^{\mu})$, $Tr (\Delta_\mu) D^{\mu} \vartheta$
and $Tr(\Gamma^{\mu} - g V^{\mu}) Tr(\Gamma_{\mu} - g V_{\mu})$. Clearly,
such terms must be included in any $U(3)_L\bigotimes U(3)_R$ symmetry based
effective field theory, where the traces of $\Delta_\mu$ and $\Gamma_\mu -
g V_\mu$ do not vanish. In addition we have introduced the coefficient
functions $W_i(X)$ which becomes  constants in the $SU(3)$ symmetry
limit. The kinetic term of the pseudoscalar mesons as 
well as their strong and electroweak interactions with the Goldstone
fields are all included in $L_A$. As for the vector
mesons, $L_V$ incorporates all interactions 
with the pseudoscalar fields. The kinetic term 
is written explicitly as $-\frac{1}{4}Tr(V_{\mu\nu}V^{\mu\nu})$.
Similar to the lagrangian of Bando et al. \cite{bando},   
the sum $L_A + aL_V$ contains, amongst other contributions,      
a vector meson mass term $\sim V_\mu V^\mu$, a vector-photon 
conversion factor $\sim V_\mu A^\mu$ and the coupling of pseudoscalar 
pairs to both vectors and photons. The latter coupling can be eliminated
by choosing a value $a=2$, which allows incorporating the conventional 
vector-dominance in the electromagnetic form-factors of
pseudoscalar mesons\cite{bando}, and eliminating the coupling of a
pseudoscalar particle to two photons. 
Data analysis \cite{benayoun98} with variants of BKY lagrangian
argue for a value $a=2.4$. This parameter is determined
anew\cite{gedalin012}  using our lagrangian.

\subsection{$SU(3)_F$ symmetry breaking}¥

As already indicated in the introduction, the expansion of
the QCD generating functional in powers of quark mass term results with 
SU(3) flavor symmetry breaking terms. There is no unique
way to introduce such terms into the lagrangian, 
but it is quite natural choosing them to be (i) Hermitian, (ii) 
proportional to powers of the quark mass matrix and, 
(iii) recover the unbroken lagrangian smoothly in the limit of vanishing 
symmetry breaking parameters. To be consistent with QCD, we add a symmetry
breaking companions to each of the unbroken lagrangian terms, using the a
procedure as described below. First, 
the Goldstone meson mass degeneracy is removed by adding
a $U(3)_L\bigotimes U(3)_R$ symmetry violating mass term.
Most generally such a term reads,
 \cite{gasser85,leut96,herera97},
\begin{equation}
	L_m = -W_0(X) + W_2(X)Tr \chi_+ +iW_3(X) Tr\chi_-~,
\label{maslag}
\end{equation}
with,
\begin{eqnarray}
	 \chi_\pm = 2B_0(\xi{\cal M^\dagger}\xi
	 \pm \xi^\dagger {\cal M} \xi^\dagger)~,
      && \qquad B_0 = m_\pi^2 / (m_u +m_d)~,
\end{eqnarray}
where $m_u, m_d, m_s$ are the up down and strange quark masses, and 
${\cal M} = diag (m_u, m_d, m_s)$.
Next, in order to incorporate symmetry  breaking corresponding to 
each of the unbroken $L_A$ and $L_V$ lagrangian densities, 
we define a universal Hermitian matrix $B$,
\begin{equation}
	B \equiv \frac{1}{4B_0(m_u + m_d + m_s)}\chi_+~.
	\label{breakm}
\end{equation}
Then, symmetry breaking companions for 
$L_A$ are constructed in two alternative ways. The first
(hereafter referred to as $Alternative ~I$) breaks octet symmetry 
($SU(3)_F$) only, 
and corresponds to a quadratic
form of the Goldstone meson kinetic energy term.  Let,
\begin{equation}
	U_8 \equiv \xi^2_8 \equiv \exp(i \frac{\sqrt{2}}{F_8}P)
	\label{u8}
\end{equation}
be the pure octet field matrix and let,
\begin{equation}
	 \bar{\Delta}_\mu = 
	 \frac{i}{2}\left\{\xi^\dagger_8 ,\partial_{\mu}\xi_8\right\}
 +\frac{1}{2}\left(\xi^\dagger_8 r_\mu\xi_8 - 
	  \xi_8 l_\mu\xi^\dagger_8\right) ~,
	\label{bdelta} 
\end{equation}
be the  octet axial vector covariant.  Then a general symmetry 
breaking lagrangian $\bar{L}_A$ would be,
\begin{eqnarray}
	 & &\bar{L}_A =
	 W_1(X)\left(c_A Tr (B\bar{\Delta}_\mu
	 \bar{\Delta}^\mu) + d_A Tr(B \bar{\Delta}_\mu
	 B\bar{\Delta}^\mu)\right)+ \nonumber \\
	 && W_4(X)d_A Tr(B\bar{\Delta}_\mu) Tr(B\bar{\Delta}^\mu)+
            W_5(X)c_A Tr(B \bar{\Delta}_\mu)D^\mu \vartheta ~. 
	\label{labreak} 
\end{eqnarray}
The second alternative ($Alternative~II$) breaks $U(3)_F$ symmetry and 
uses the nonet axial vector
$\Delta^\mu$ ( instead of $\bar{\Delta}^\mu$ as above).
Namely, 
 \begin{eqnarray}
     &&\bar{L}_{A}=
     W_{1}(X)\left(c_ATr(B\Delta_\mu\Delta^\mu)
         + d_ATr(B\Delta_\mu B\Delta^\mu)\right)
      \nonumber \\
         &&+W_{4}(X)\left(c_ATr(B\Delta_\mu)Tr(\Delta^\mu) +
         d_ATr(B\Delta_\mu)Tr(B\Delta^\mu)\right) +
         \nonumber\\
         && W_{5}(X)c_ATr(B\Delta_\mu)D^\mu \vartheta ~.
        \label{labreak1}
        \end{eqnarray}
This expression gives a bilinear meson kinetic energy term. Similarly the 
asymmetric companion of ${L}_V$ would be,
 \begin{eqnarray}
  &  & \bar{L}_V =
	 \tilde{W}_1(X)\left(c_V Tr (B[\Gamma_\mu -gV_\mu] 
	[\Gamma^\mu -gV^\mu])\right. +
	\nonumber\\
	&& \left. d_V Tr (B[\Gamma_\mu-gV_\mu] B [\Gamma^\mu -gV^\mu])\right) +
	 \nonumber\\
	 &&\tilde{W}_4(X)\left(c_VTr(\Gamma_\mu - gV_\mu)Tr(B[\Gamma^\mu -
           gV^\mu])\right.
	 \nonumber\\
	&& +\left. d_VTr(B[\Gamma_\mu - gV_\mu])Tr(B[\Gamma^\mu -
         gV^\mu])\right) ~.    
	\label{lvbreak}
 \end{eqnarray}	
In the expressions above, $c_A, d_A, c_V~\mbox{and}~d_V$ are 
symmetry breaking  scales to be determined from 
data analyses. It is to be stressed 
that $\bar{L}_A$ and $\bar{L}_V$ differ also from the ones defined
in Refs.\cite{bando,bramon95,benayoun981}. First, like
our symmetric $L_A$ and $L_V$, the asymmetric companions $\bar{L}_A$ and
$\bar{L}_V$ involve additional terms which are absent in the $SU(3)$
limit. Secondly, in the original work of Bando et al.\cite{bando}, and
variants of this model\cite{benayoun981}, the scales $d_i$ are taken
to be equal to $(c_i/2)^2$, based on their assumption that the physical
fields are related to the bare fields 
through rescaling, i.e.,
\begin{equation}
\phi_{ph} =  \lambda^{1/2} \phi \lambda^{1/2}~.
\end{equation}  
Contrary to this assumption, Bramon et al.\cite{bramon95} have assumed
$d_i = 0$. In the our model, the field rescaling is dictated by the
diagonalization procedure and is uniquely determined by the form of the 
lagrangian\cite{gedalin01}.
For the more general lagrangian constructed above, 
and in order to reproduce the actual spectra of the physical
mesons\cite{gedalin012} the constants $c_i$ and $d_i$ must then be treated as
independent parameters. 
Thirdly, unlike Bando et al.\cite{bando} the symmetry breaking 
matrix $B$ is Hermitian. It is similar (but not identical) to that of  
Bramon et al.\cite{bramon95} and  rather close to the one 
proposed by Benayoun and O'Connell\cite{benayoun981}. Albeit,
the lagrangian  is constructed in close analogy with
QCD and allows for symmetry breaking in a universal manner.
Particularly, the matrix
$B$, Eqn. \ref{breakm}, enables us to maintain the 
QCD ratios of isospin to $SU(3)$ symmetry breaking  scales.

Summing all terms, including the well known Wess-Zumino-Witten term
$L_{WZW}$ \cite{witten83,callan84}, the lagrangian assumes the form,
\begin{equation}
  L = L_A + \bar{L}_A + L_m + a(L_V + \bar{L}_V) - 
  \frac{1}{4}Tr(V_{\mu\nu}V^{\mu\nu}) + L_{WZW} + \ldots ~,
  \label{elag}
\end{equation}
where "$\ldots$"  stands for terms accounting for the
regularization of one loop contributions 
\cite{bijnens90,bijnens901,gasser85}. Note that ${\bar L}_V$
includes vector meson mass terms which also violate symmetry.
Generally speaking, a symmetry breaking companion for the vector meson
kinetic term $-\frac{1}{4}Tr(V_{\mu\nu}V^{\mu\nu})$ should have
been added also.  However, at present, no evidence exists for such
asymmetric terms and therefore will not be considered in the 
following discussion. 
The lagrangian constructed above should describe
the mass splitting  of the pseudoscalar and vector mesons (see for example
\cite{pdg00}).
Since the ground state of the field $U$ is
proportional to the unit matrix, one can set the auxiliary field 
$\vartheta = 0$ \cite{gasser85}. With these simplifying
assumptions  the quantities
$W_i$ and $\tilde{W}_i$ become functions of the singlet field ($\eta_0$)
only. 
To lowest order their expansions read,  
 \begin{eqnarray}
 	 &  & W_{0} = const + F^4_8\left( w_0 \frac{\eta^2_0}{F^2_0} 
 	 + \ldots\right)~, 
 	\label{wco} \\
 	 &  & W_{1} = F^2_8\left(1 + w_1 \frac{\eta^2_0}{F^2_0} 
 	 + \ldots\right)~,
 	\label{wc1} \\
 	 &  &  W_{2} = \frac{F^2_8}{4}\left(1 + w_2 \frac{\eta^2_0}{F^2_0} 
 	 + \ldots\right)~, 
 	\label{wc2} \\
 	 &  &  W_{3} = \frac{ F^2_8}{2}(w_3 \frac{\eta_0}{F_0} +  \ldots)~,
 	\label{wc3} \\
 	&  & W_{4} = \frac{F^2_0 - F^2_8}{3}\left(1 + 
 	 w_4 \frac{\eta^2_0}{F^2_0} +  \ldots\right)
 	\label{wc5} \\
 	 &  & W_{5} = F^2_8\left(\bar{w}_5 + 
 	 w_5 \frac{\eta^2_0}{F^2_0} +  \ldots\right)
 	\label{wc5} \\
 	 &  & W_{6} =  F^2_8\left(\bar{w}_6 + 
 	 w_6 \frac{\eta^2_0}{F^2_0} +  \ldots\right)~, 
 	\label{wc6} \\
 	 &  & \tilde{W}_{1} = F^2_8\left(1 + \tilde{w}_1 \frac{\eta^2_0}{F^2_0}  
 	 + \ldots\right)~, 
 	 \label{wwc1} \\
 	 &&\tilde{W}_{4} = F^2_8\left(\tilde{w}_4 + \tilde{w}_5
         \frac{\eta^2_0}{F^2_0} + \ldots\right)~. 
 	\label{wwc4}
 \end{eqnarray}
Here the coefficients $w_i,\bar{w}_i ~~\mbox{and}~~ \tilde{w}_i$ 
are free parameters not yet determined. In the applications to be
discussed below, however, only a few combinations (rather than each) of
these parameters are needed.

\section{Symmetry breaking and pseudoscalar meson mixing}
  
Symmetry breaking terms affects the form of the kinetic and mass lagrangian 
densities of the Goldstone mesons which in turn uniquely determine  
the state and decay constants mixing schemes\cite{gedalin01}. 
In what follows we realize the mixing schemes, corresponding to the
$Alternative~ I$ ( Eqn. \ref{labreak} ) and  the 
$Alternative~ II$ ( Eqn. \ref{labreak1}) lagrangians. 
These two alternatives allow us to evaluate 
$SU(3)_F ~versus ~U(3)_F$ symmetry
breaking. A third scheme corresponding to the QFB of
Feldmann et al.\cite{feld98},  is also realized from the 
$Alternative~~II$ by requiring nonet symmetry ($F_0 = F_8$). 

Generally, for
any EFT the kinetic and mass terms can be written in the form, 
\begin{equation}
	L_{km} = 
	\frac{1}{2}(\partial_\mu \Phi){\cal K}(\partial^\mu \Phi) + 
	\frac{1}{2}\Phi {\cal M}^2\Phi~,
	\label{lagkm}
\end{equation}
where $\Phi$ represents  the intrinsic meson field matrix,  
${\cal K}~\mbox{and}~{\cal M}^2$ are 
the kinetic and mass matrices. In the presence of symmetry 
breaking ${\cal K}~\mbox{and}~{\cal M}^2$ are non-diagonal, and
Eqn. \ref{lagkm}
does not have the standard quadratic form as invoked by the
Klein-Gordon equation for the physical fields. 
It can be reduced into this standard form by applying 
three consecutive steps, which transform the intrinsic fields into the
physical meson fields according to \cite{gedalin01},
\begin{equation}
        \Phi =  \Theta \Phi_{ph}~;~~~ \Theta = \Upsilon R \Omega~.
        \label{thetatr1}
\end{equation}
Here $\Upsilon$ represents a unitary transformation which diagonalize the
kinetic matrix ${\cal K}$, $R$ stands for rescaling of the fields, and $
\Omega$
is another unitary matrix which diagonalize the resulting mass matrix.
Let us work in details a general scheme for the mixing of
two states, where presumably, the $\eta$ and $\eta '$ are 
linear combinations of an octet ($\eta_8$) and 
a singlet ($\eta_0$), only.
Using the symmetry breaking $Alternative~II$ lagrangian,
Eqn.\ref{labreak1}, 
the kinetic and mass terms are bilinear and the corresponding matrices 
read,
\begin{equation}
	 {\cal K} = \left(
	\begin{array}{cccc}
		\kappa_\pi & 0 & 0 & 0  \\
		0 & \kappa_K & 0 & 0  \\
		0 & 0 & \kappa_{88} & \kappa_{08}  \\
		0 & 0 &\kappa_{08} & \kappa_{00}
	\end{array}\right)~,\qquad 
	 {\cal M}^2 = \left(
	\begin{array}{cccc}
		\mu^2_\pi & 0 & 0 & 0  \\
		0 & \mu^2_K & 0 & 0  \\
		0 & 0 & \mu^2_{88} & \mu^2_{08}  \\
		0 & 0 & \mu^2_{08} & \mu^2_{00}
	\end{array}\right)~.
	\label{¥}
\end{equation}
¥Here,   
\begin{eqnarray}
     &  & \kappa_\pi = 1 ~,
  \nonumber\\
  	 &  & \kappa_K = 1 + \frac{1}{2}c_A~,
  	\nonumber \\
    & & \kappa_{88} =
    1+\frac{2}{3}\left(c_A + \frac{2r^2+1}{3r^2} d_A\right)~,
	 \nonumber\\
         & ¥ &  \kappa_{08} 
    =-\frac{r\sqrt{2}}{6}\left(\frac{1+r^2}{r}c_A+
    2\frac{2r^2+1}{3r}d_A\right)~,
        \nonumber \\
   ¥ & ¥ &  \kappa_{00} = 1+\frac{1+5r^2}{18}c_A +\frac{1+r^2}{6}d_A~,
         \nonumber \\
    &  & \mu^2_\pi = m^2_\pi = 2mB_0 ~,
	\nonumber \\
    &  & \mu^2_K = \frac{m^2_\pi}{2}(1+\frac{m_s}{m})~,
	\nonumber \\
    & ¥& \mu^2_{88} = \frac{m^2_\pi}{3}(1+2\frac{m_s}{m})~,
	\nonumber \\
    & ¥ &\mu^2_{08} = \frac{\sqrt{2}}{3}
	 m^2_\pi (1-\frac{m_s}{m})(1 + \sqrt{6}w_3)r~, 
	\nonumber \\
    & ¥& \mu^2_{00} = \frac{m^2_\pi}{3}(2+\frac{m_s}{m})\left[1+
        2\sqrt{6}w_3 -3w_2+\frac{6mF^2_8}{m^2_\pi(2m+m_s)}w_0\right]r^2~,
        \label{kmmat}
\end{eqnarray}¥
and $~r= F_8/F_0$ is a measure of nonet symmetry breaking. For simplicity ,
small terms of order $\sim  c_A m/m_s,~ d_A m/m_s~$ are neglected and 
isospin symmetry ($m_u =m_d=m$) is assumed in the expressions listed
above. 
Note that $\cal K$ and ${\cal M}^2$ have non-diagonal $2\times 2$
submatrices, and as demonstrated in Ref. \cite{gedalin01} the transformation
$\Theta$ involves   
two mixing angles and two rescaling parameters.
First as in  Ref. \cite{gedalin01} we diagonalize 
${\cal K}$ using the unitary transformation,
         \begin{equation}
         \left(\begin{array}{c}
                        \pi \\
                        K     \\
                        \eta_8  \\
                        \eta_0
                \end{array}\right) =
                 = \Upsilon\left(\begin{array}{c}
                        \pi \\
                        K    \\
                        \bar{\eta}_8 \\
                        \bar{\eta}_0
        \end{array}\right)~, \quad \Upsilon = \left(\begin{array}{cccc}
                 1 & 0 & 0 & 0\\
                 0 & 1 & 0 & 0\\ 
                 0 & 0 & \cos\lambda & \sin\lambda  \\
                 0 & 0 & -\sin\lambda & \cos\lambda
                \end{array}\right)~.
         \label{upstr}     
         \end{equation}  
This leads to,
\begin{equation}
    L_{km}  =
            (\partial_\mu\pi)^2 +    
            \kappa_K(\partial_\mu K)^2 + \kappa_0
                (\partial_\mu\bar{\eta}_0)^2  + 
                 (\bar{\eta}_8,\bar{\eta}_0)
                \Upsilon^{-1}{\cal M}^2\Upsilon
                \left(\begin{array}{c}
                        \bar{\eta}_8  \\
                        \bar{\eta}_0
                \end{array}\right)~,
         \end{equation}
with,
\begin{eqnarray}
	 &  & \kappa_8 = \frac{1}{2}\left[\kappa_{88} + 
	 \kappa_{00} - \sqrt{(\kappa_{00} - \kappa_{88})^2 + 
	 4\kappa^4_{80}}\right]~,
	\label{¥} \\
	 &  & \kappa_0 = \frac{1}{2}\left[\kappa_{88} + 
	 \kappa_{00} + \sqrt{(\kappa_{00} - \kappa_{88})^2 + 
	 4\kappa^4_{80}}\right]~,
	\label{¥}
\end{eqnarray}¥
and,
\begin{equation}
    \frac{\tan \lambda}{1-\tan^2\lambda}
    =  \frac{\kappa_{80}}{\kappa_{00} - \kappa_{88}}~.
	\label{¥}
\end{equation}¥
Next we rescale the fields ($\pi, K, \bar{\eta}_8, \bar{\eta}_0$)
into ($\pi, \hat{K}, \hat{\eta}_8, \hat{\eta}_0$) using,
\begin{equation}
R = diag(1,~z_K,~z,~f) = 
diag(1, \frac{1}{\sqrt{\kappa_K}},~\frac{1}{\sqrt{\kappa_8}},
~\frac{1}{\sqrt{\kappa_0}})~.
\label{rtran}¥
\end{equation}
Following these two steps, the  kinetic term acquires the standard
quadratic form and the resulting mass matrix is,
\begin{equation}   
    \tilde{{\cal M}}^2 = R\Upsilon^{-1}{\cal M}^2 \Upsilon R = 
    \left(
    \begin{array}{cccc}
    \tilde{\mu}^2_\pi & 0 & 0 & 0\\
    0        &\tilde{\mu}^2_K & 0 & 0 \\
    0 & 0 &	\tilde{\mu}^2_{88} &  \tilde{\mu}^2_{80} \\
    0 & 0 &	\tilde{\mu}^2_{80} &  \tilde{\mu}^2_{00}
    \end{array}¥
     \right)~,
    \label{resmassm}
\end{equation}
with,
\begin{eqnarray}
     &   & \tilde{\mu}^2_{\pi} = \mu^2_{\pi}~,\\
     &   & \tilde{\mu}^2_K = \mu^2_K z^2_K~,
     \label{massk} \\
     &   & \tilde{\mu}^2_{88} = 
	 z^2 \left(\cos^2 \lambda \mu^2_{88} + \sin^2 \lambda 
	 \mu^2_{00} -2\sin\lambda\cos\lambda \mu^2_{80}\right)~,
	\label{¥} \\
     &   & \tilde{\mu}^2_{80} = 
	zf\left(\sin\lambda\cos\lambda (\mu^2_{88}-\mu^2_{00}) + 
        (\cos^2\lambda - \sin^2\lambda)\mu^2_{80}\right)~,
    \label{¥} \\
     &   & \tilde{\mu}^2_{00} = f^2\left(\sin^2\lambda \mu^2_{88} +
    \cos^2\lambda \mu^2_{00} + 2\sin\lambda\cos\lambda \mu^2_{80}\right)~.
    \label{}
\end{eqnarray}
As a last step we diagonalize the mass matrix, Eqn. \ref{resmassm}, by
applying a second
unitary transformation, 
\begin{equation}
        \left(  
        \begin{array}{c}
                \hat{\eta}_8  \\
                \hat{\eta}_0
        \end{array}
        \right) =\Omega \left(
        \begin{array}{c} 
                \eta  \\
                \eta '
        \end{array}
        \right)~, \quad \Omega = \left(
        \begin{array}{cc}
                \cos\chi & \sin\chi  \\
                -\sin\chi & \cos\chi
        \end{array}
        \right) ~,
        \label{omegatr}
\end{equation} 
and take the eigenvalues to be equal to the masses of the physical
particles, i.e.,
\begin{eqnarray}
     &  & m^2_\pi = \mu^2_\pi~, \\
     &  & m^2_K = \mu^2_K z^2_K ~,\label{kmass}\\
     &  & m^2_{\eta} = \frac{1}{2} \left( \tilde{\mu}^2_{88} +
    \tilde{\mu}^2_{00} - \sqrt{ (\tilde{\mu}^2_{88} -
    \tilde{\mu}^2_{00})^2 + 4(\tilde{\mu}^2_{80})^2 } \right)~,
    \label{} \\
     &  &m^2_{\eta '} = \frac {1}{2}\left( \tilde{\mu}^2_{88} +
    \tilde{\mu}^2_{00} + \sqrt{(\tilde{\mu}^2_{88} - 
    \tilde{\mu}^2_{00})^2 + 4(\tilde{\mu}^2_{80})^2}\right)~,
    \label{etamass}
\end{eqnarray}
and,
\begin{equation}
    \frac{\tan \chi}{1-\tan^2\chi} 
    = - \frac {\tilde{\mu}^2_{80}}
    { \tilde{\mu}^2_{88}- \tilde{\mu}^2_{00} }~.
    \label{¥}
\end{equation}¥
By definition the physical $\eta $ and $\eta '$ fields
are orthogonal and are related to the intrinsic fields via,
\begin{equation}
        \left(  
        \begin{array}{c}
                \eta_8  \\
                \eta_0
        \end{array}
        \right) = \Theta '\left(
        \begin{array}{c} 
                \eta  \\
                \eta'
        \end{array}
        \right)~,
        \label{}   
\end{equation}
where,
\begin{equation}
     \Theta ' =         \left(
        \begin{array}{cc}
    z\cos\lambda\cos\chi-f\sin\lambda\sin\chi &
      z\cos\lambda\sin\chi+f\sin\lambda\cos\chi  \\
         -z\sin\lambda\cos\chi-f\cos\lambda\sin\chi &
         -z\sin\lambda\sin\chi+f\cos\lambda\cos\chi
        \end{array}
        \right)~.
        \label{tetatr}
\end{equation} 
Using this transformation, we may rewrite the nonlinear representation,
Eqn. \ref{pmfield} in terms of the physical fields, e.g., 
 \begin{equation}
 	U = \exp{i\frac{\sqrt{2}}{F_\pi}{\cal P}}~,
 	\label{ufield}
 \end{equation}
 where $\cal P$ stands for the pseudoscalar nonet matrix,
  \begin{equation}
{\cal P} = \left(\begin{array}{ccc}
		\frac{\pi^0}{\sqrt{2}}+\frac{1}{\sqrt{6}}(X_\eta \eta +
		X_{\eta'}\eta') & \pi^+ &{z}_sK^+  \\
		\pi^- & - \frac{\pi^0}{\sqrt{2}}+ \frac{1}
		{\sqrt{6}}(X_\eta \eta +X_{\eta'}\eta')  &{z}_sK^0  \\
		{z}_sK^- & {z}_s\bar{K}^0 &\frac{1}{\sqrt{6}}
		(Y_\eta \eta + Y_{\eta'}\eta')
	\end{array}\right)~.
	\label{pnonet}
	\end{equation}
The coefficients $X_i$ and $Y_i, (i = \eta, \eta ')$ are listed in Table
\ref{xyfac}. It is worthy to mention that by taking 
the eigenvalues of the resulting mass matrix to be equal to the masses of
the physical mesons we have eliminated explicit dependence on the model
free parameters $w_i$. Consequently, the
mixing angles $\lambda~\mbox{and}~\chi$, the rescaling parameters
$z~\mbox{and}~ f$, 
the coefficients  $X_i$ and $Y_i$, and the pseudoscalar nonet matrix 
${\cal P}$ become functions of just three symmetry
breaking scales $c_A, d_A~\mbox{ and}~ r$. 

This enables us to incorporate symmetry breaking $indirectly$
in the pseudoscalar nonet matrix, Eqn. \ref{pnonet}.  We refer to 
these  as $indirect$ as opposed to {\it direct\/} symmetry breaking 
due to the broken $\bar{L}_A$ and $\bar{L}_V$ lagrangian companions. 

At this stage some comments are in order : \\
(i) A better known expression of the transformation  $\Theta$ 
is ,
\begin{equation}
        \Theta  = 
    \left(\begin{array}{cc}
    \cos\theta_{\eta'} & ~\sin\theta_{\eta}  \\  
         -\sin\theta_{\eta'} & ~\cos\theta_{\eta}
        \end{array}\right)diag(z_1, z_2)~,
        \label{tetatr1}
\end{equation}  
Where,
\begin{eqnarray}
	z^2_1 = z^2 \cos^2\chi +f^2\sin^2\chi,~&~z\cos\chi = 
	z_1\cos\psi_1,~& ~f\sin\chi = - z_1\sin\psi_1,
	\label{¥} \\
	z^2_2 = f^2 \cos^2\chi +z^2\sin^2\chi,~&~f\cos\chi = 
        z_2\cos\psi_2,~ & ~z\sin\chi = - z_2\sin\psi_2,
	\label{¥} \\
	\tan\psi_1 = - \frac{f}{z}\tan\chi, ~ & ~	
        & ~
        \tan\psi_1 =   \frac{f^2}{z^2}\tan\psi_2,  
	\label{zpsi} \\                              
                       \theta_\eta = \lambda-\psi_2~,~ & ~
                       \qquad \theta_{\eta'} = \lambda-\psi_1~.
        \label{}  
\end{eqnarray}
This exact general expression of the field transformation involves two mixing 
angles $\theta_\eta ~\mbox{and}~ \theta_{\eta'}$ as well as two field rescaling
parameters $z_1 ~\mbox{and}~z_2$, and is dictated by the kinetic and mass
terms of the lagrangian presumed. A similar two angle scheme was 
proposed by Escribano and Fr\`{e}re \cite{escrib99,kaiser98}, however their
transformation does not account for the field rescaling and violates orthogonality 
of the $\eta$ and $\eta '$ states.

 (ii) With the $Alternative ~I$ symmetry breaking lagrangian ${\bar L}_A$,
\ref{labreak}, the kinetic term  assumes a quadratic form,  with
$\kappa_{88}=1+2(c_A+d_A)/3,~\kappa_{08}=0,~\mbox{and}~\kappa_{00}=1$. All
other matrix elements of ${\cal K}$ and ${\cal M}^2$ remain the same as in
Eqn. \ref{kmmat}. Consequently, $z=1/\sqrt{1+2(c_A+d_A)/3}$ and $f=1$, and
the angle $\lambda$ vanishes,   so that the
$Alternative ~I$ scheme involves a single mixing angle.

(iii) In the limit of exact nonet symmetry, $r=1~(F_8=F_0)$, the
$Alternative~II$ pseudoscalar meson kinetic matrix becomes diagonal in
the so called quark flavor basis (QFB), where the flavorless mesons 
are represented by fields with or without strange quark content 
($\eta_s \sim s \bar{s}$ and $\eta_q \sim (u \bar{u} + d \bar{d})/\sqrt 2$). 
With this basis the kinetic matrix is diagonal, which leads to
a one mixing angle scheme. Note that  Feldmann et
al.\cite{feld98,feld98a,feld98b} do not account for field rescaling and
therefore, their mixing angle $\phi$ is equivalent to our mixing angle
$\chi$, which defines the transformation of the rescaled fields into the
physical ones.

(iv) As a result from the diagonalization procedure, in all mixing
schemes, the rescaling parameters and mixing angles are functions of the
symmetry breaking scales and can not be taken to be free parameters, as
commonly assumed in previous studies
\cite{benayoun981,benayoun98,gedalin00}.

To conclude this section, we briefly consider the vector meson mixing
problem. Here, the kinetic energy term has the standard quadratic form. 
Thus one needs diagonalizing the mass matrix only, which
effectively depends on four parameters
$m^2_V=2F^2_8ag,~c_V,~d_V~\mbox{and}~\tilde{w}_4 $, where 
$m_V$ stands for  the vector meson nonet symmetric mass. All four 
parameters can be fixed by taking the calculated physical meson masses 
to be equal to their experimental values. Following a similar diagonalization 
procedure as above, it is straightforward to show that the physical vector 
nonet  matrix with non-ideal mixing reads,
\begin{equation}
     V = \left(\begin{array}{ccc}
		\frac{1}{\sqrt{2}}(\rho^0+\omega+\epsilon\phi)
		& \rho^+ & K^{*+}  \\
		\rho^- &
              \frac{1}{\sqrt{2}}(-\rho^0+\omega+\epsilon\phi) 
		  & K^{*0}  \\
		K^{*-} & \bar{K}^{*0} &\phi -\epsilon \omega
	\end{array}\right)~.
	\label{vnonet}
	\end{equation}   
with $\epsilon = 0.047 \pm 0.001$, being a measure of the non-strange (strange) 
admixture in $\phi ~(\omega)$. Nearly the same value of $\epsilon$ is deduced 
from the width of the $\phi \to \pi \gamma$ decay (see next section). 
In practice, the vector meson mixing angle turns to be very close to the 
one of ideal mixing, and likewise, the corresponding strange (non-strange) 
admixtures in the nonet vector matrix are small. Since we worked in 
the exact isospin symmetry limit, the vector field, Eqn. \ref{vnonet}, does 
not account for $\rho - \omega$ mixing. The derivation of the broken 
matrix is straightforward and will be given elsewhere\cite{gedalin012}.

\bigskip

\section{ Radiative Decay Widths}

The main objective in the following section  is to apply our theory to
calculate decay width of the anomalous processes listed in table
\ref{vpgpred1}. Details of the calculations are described in previous
publications and will not be repeated ( see for example
Ref.\cite{benayoun99}).  Our analyses generalize the treatments of 
Refs.\cite{bijnens90,bramon95,benayoun98,benayoun99}, in the sense that 
we treat all nonet mesons on equal footing and most importantly we use
mixing schemes which are uniquely determined by the lagrangians presumed.

In our analyses we use pseudoscalar and vector field matrices expressed in 
terms of physical fields and therefore allow for $indirect$ and $direct$
symmetry breaking effects. The lagrangian is factorized in the form,
\begin{eqnarray}
         &  &  L_{P\gamma\gamma} =  L_{P\gamma\gamma} +
          c_W  \bar{L}_{P\gamma\gamma},
         \\
         &  &  L_{VP\gamma} =  L_{VP\gamma} +
          c_W \bar{L}_{VP\gamma}~,
         \label{dlag}
\end{eqnarray}
where $L_{P\gamma\gamma}$ ($L_{VP\gamma}$) represents the
overall contribution
of the unbroken anomalous lagrangian to the $P\gamma \gamma$
($VP\gamma$) interaction, and $\bar{L}_{P\gamma\gamma}$
($\bar{L}_{VP\gamma}$) is the corresponding direct symmetry 
breaking companion.  Here $c_W$ stands for the direct symmetry 
breaking parameter. As mentioned above, both
$L_{P\gamma\gamma}$ and $L_{PVV}$ account for symmetry breaking 
via the pseudoscalar and vector meson matrices. To  write these terms 
explicitly, we start from the anomalous lagrangian\cite{bijnens90},
\begin{equation}
	L_{anomalous} =    L^{(0)}_{VVP} +
	 L_{WZW}(P\gamma\gamma)~.
	\label{dlag0}
\end{equation}
In terms of the covariants $\Delta_\mu$, $\Gamma_\mu-gV_\mu$,
$V_{\mu \nu}$ and $  \Gamma_{\mu\nu} = \partial_\mu \Gamma_\nu -
\partial_\nu \Gamma_\mu - i[\Gamma_\mu, \Gamma_\nu]$ 
(see sect II for definitions), the pseudoscalar-vector-vector (PVV)
coupling, $L^{(0)}_{VVP}$, at lowest order has at most six contributions, 
\begin{eqnarray}
	 &&  L^{(0)}_{VVP} = g_1\epsilon^{\mu\nu\alpha\beta}
	 Tr(V_{\mu\nu} 
	 [V_\alpha -\frac{1}{g}\Gamma_\alpha]\Delta_\beta) +
	 g_2\epsilon^{\mu\nu\alpha\beta}Tr(\Gamma_{\mu\nu} 
	 [V_\alpha -\frac{1}{g}\Gamma_\alpha]\Delta_\beta) +
	 \nonumber \\
	 && g_3\epsilon^{\mu\nu\alpha\beta}Tr(V_{\mu\nu})
	 Tr([V_\alpha -\frac{1}{g}\Gamma_\alpha]\Delta_\beta) +
	  g_4 \epsilon^{\mu\nu\alpha\beta}Tr(\Gamma_{\mu\nu})
	 Tr([V_\alpha -\frac{1}{g}\Gamma_\alpha]\Delta_\beta) +
	 \nonumber \\
	 && g_5\epsilon^{\mu\nu\alpha\beta}Tr(V_{\mu\nu})
	 Tr(V_\alpha -\frac{1}{g}\Gamma_\alpha )Tr(\Delta_\beta) +
	 \nonumber \\
	 && g_6\epsilon^{\mu\nu\alpha\beta}Tr(\Gamma_{\mu\nu})
	 Tr(V_\alpha -\frac{1}{g}\Gamma_\alpha )Tr(\Delta_\beta)~,
	\label{vvplag}
\end{eqnarray}
where $g_i,~(i=1,...6)$ are  arbitrary coefficient functions of the
variable X. By rearranging terms we may write $ L_{anomalous}$ in a
compact form, e.g., 
\begin{eqnarray}
	 &  & L_{VP\gamma}= 
	 g_V \frac {e}{F_\pi}\epsilon^{\mu\nu\alpha\beta}\partial_\mu 
	 A_\nu Tr( Q\{\partial_\alpha V_\beta, {\cal P}\})~, 
	\label{vpgamma}\\
	 &  &  L_{P\gamma\gamma} =
	 g_P\frac{e^2}{2F_\pi}\epsilon^{\mu\nu\alpha\beta}\partial_\mu 
	 A_\nu\partial_\alpha A_\beta Tr(\left\{ Q^2,{\cal P}\right\})~.
	\label{pgamma} 
\end{eqnarray}
Here $g_P$ and $g_V$ represent certain combinations  of the function
coefficients $g_i$ and constants in $L_{WZW}$. 
As in section II, the  direct symmetry breaking
terms are constructed by inserting the quantity B, Eqn. \ref{breakm},
in the expressions above, i.e.,
\begin{eqnarray}
     &  & {\bar L}_{VP\gamma} = 
     g_V \frac {e}{F_\pi}\epsilon^{\mu\nu\alpha\beta}\partial_\mu 
	 A_\nu Tr( Q \{B,\{\partial_\alpha {V}_\beta ,{\cal P}\}\})~,
	\label{vpgammab} \\
     &  &  {\bar L}_{P\gamma\gamma}= 
	 g_P\frac{e^2}{2F_\pi}\epsilon^{\mu\nu\alpha\beta}\partial_\mu 
	 A_\nu\partial_\alpha A_\beta Tr(\left\{ Q^2,\{B,{\cal P}\}\right\})~.
	\label{ptwogammab}
\end{eqnarray}

\subsection{The $V\to P \gamma$ and $P \to V \gamma$ Processes}
The relevant vertices for a vector (pseudoscalar) meson decaying into a 
pseudoscalar (vector) meson and a photon are,
 \begin{equation}
  {\cal {V}} (VP\gamma) = 
  - i g_V\frac{e}{F_\pi}\it{v}(VP)\epsilon^{\mu\nu\alpha\beta}k_\mu 
  e^{(\gamma)}_\nu p_\alpha e^{(V)}_\beta ~,
 \label{vpgamma} 
 \end{equation}
 where $ e^{(V)}_\nu$ ($p$) and $ e^{(\gamma)}_\nu$ ($k$) are 
 the polarization (four-momentum) of the vector meson and final photon,
 respectively. The quantities $\it{v}$ incorporate all internal
symmetries of the processes under discussion and are listed in Table
\ref{vfac}. In terms of these vertices the widths of the decays 
$V \rightarrow P \gamma$ and $P \rightarrow V \gamma$ are,
\begin{eqnarray}
 	 &  & \Gamma (VP\gamma) = G_V\frac{(m^2_V - m^2_P)^3}{m^3_V F^2_8}
 	|\it{v}(VP)|^2~,
 \label{gvp}	 \\
     &  & \Gamma (P V\gamma) =3 G_V\frac{(m^2_{P} - m^2_V)^3}
     {m^3_{P} F^2_8}|\it{v}(V P)|^2~,
     \label{pvg}	
 \end{eqnarray}
 with, 
 \begin{equation}
 	G_V = \frac{e^2}{4\pi}\frac{g^2_V}{24}~.
 \end{equation}

\subsection{The $P\rightarrow \gamma\gamma$ decays}
The vertices for decays of a pseudoscalar meson into two photons are,
 \begin{equation}
    {\cal V} (P\gamma\gamma) =  - 2i g_P
    \frac{e^2}{F_8}\bar{\it{v}}(P)\epsilon^{\mu\nu\alpha\beta}k_{1\mu} 
    e^{(\gamma)}_\nu k_{2\alpha} e^{(\gamma)}_\beta ~,
\label{ptwog} 
\end{equation}
 with $ e^{(\gamma)}_\nu$ and $ e^{(\gamma)}_\alpha$ being
 the polarizations of the final photons, and $k_1$, $k_2$ their 
 corresponding four-momenta. Again, the functions $\bar{\it{v}}(P)$  
contain the internal symmetry information and are listed in Table
\ref{vfac}. 
With these vertices the decay rate is given by,
 \begin{equation}
  \Gamma (P\gamma\gamma) = G_P\frac{ m^3_P}{ F^2_8}|\bar{v}(P)|^2~,
  \label{p2gamma}¥
 \end{equation}
 where,
 \begin{equation}
 	G_P = \frac{\pi}{2}\left(\frac{e^2}{4\pi}\right)^2\left(\frac{g_P
}
 	{9}\right)^2~.
 \end{equation}
 
\subsection{Numerical Analysis and Results} 

The expressions quoted above for the decay widths 
involve three symmetry breaking scales ($c_A,~d_A,~r$),  
a direct symmetry breaking scale ($c_W$), and two coupling constants
($g_V,~g_P$). We fix the coupling constants from experiment and we
treat the other quantities as free model parameters to be determined from
global fit. As a general rule the experimental masses 
and decay widths used are the best fit values reported in the latest
review of particle properties\cite{pdg00}. The pion radiative constants is
taken to be $F_\pi = F_8 = 93 MeV$.

Consider now the values of the coupling constants.
From the experimental width of the $\omega \rightarrow \pi
\gamma$ decay, and  Eqn. \ref{gvp}, one obtains
$\Gamma(\omega\pi\gamma) = G_V
(m^2_\omega - m^2_\pi)^3/ (m^3_\omega F^2_8) = (716 \pm 43) KeV$\cite{pdg00}. 
This relation yields, 
\begin{eqnarray}
 G_V = (1.44\pm0.04)\cdot10^{-5}~, &\qquad &
        g_V = 0.220 \pm 0.006~.
        \label{gvec}
   \end{eqnarray}
In practice, these same constants  are obtained from  a similar relation,
$\Gamma(\rho\pi\gamma) = G (m^2_\rho - m^2_\pi)^3/
(m^3_\rho F^2_8) = (76 \pm 10) KeV$\cite{pdg00} for  
the $\rho \rightarrow \pi \gamma$ decay.  
Similarly, from the decay width of $\pi \to \gamma \gamma$ one
finds,
       $ \Gamma(\pi^0\gamma\gamma) = 9 G_P m^3_\pi/2F^2_8 
        = (7.8 \pm 0.55) eV$\cite{pdg00}, 
\begin{eqnarray}
G_P=(4.91\pm0.07)\cdot10^{-8}, &\qquad &
        g_P = 0.073 \pm 0.001~.
        \label{gps}
   \end{eqnarray}
Furthermore, the vector meson strange-nonstrange admixture parameter 
$\epsilon$ can be deduced from the decay widths of the $\phi$ and 
$\omega$ mesons into a pion and a photon. The $\phi\rightarrow \pi\gamma$ 
decay is forbidden in the limit of ideal mixing. However, from 
Eqns. \ref{gvp} and the experimental decay rates\cite{pdg00} one obtains,  
 \begin{equation}
          \frac{\Gamma(\phi\pi^0\gamma)}{\Gamma(\omega\pi^0\gamma)} =
        2\epsilon^2 \left[\frac{(m^2_\phi - m^2_\pi)m_\omega}
         {(m^2_\omega - m^2_\pi)m_\phi}\right]^3 = 
        \frac{(5.8 \pm 0.6)KeV}{(716\pm 43) KeV} = 0.008 \pm 0.001~.
\end{equation}
This ratio gives $\epsilon = 0.043 \pm 0.004~$, a value which is 
equal within experimental error, to that deduced from vector mass matrix 
diagonalization(see section II).  The calculated decay widths and parameter
values as obtained from global fit to data using different mixing schemes 
are summarized in Tables \ref{fit} and \ref{vpgpred1}. Based on the 
$\chi^2/dof$ values listed, the one mixing angle $Alternative ~I$ 
scheme provides the best explanation of the data. This observation 
remains valid, should we have used an older set of data\cite{pdg98}, 
though the resulting fit qualities are slightly poorer.

\underline{{\bf $Alternative~ I$~~-One mixing angle scheme}}~~
First we recall that in this case $f=1$,  $\kappa_{08} = 0$ and 
$\lambda = 0$. This simplifies Eqns.\ref{upstr},\ref{rtran},\ref{omegatr},
 \ref{tetatr} and allows expressing the mixing angle in terms of 
the physical meson masses, 
\begin{equation}
     \frac{\tan \theta_P}{1-\tan^2\theta_P}
     = -\sqrt{\left(\frac{m^2_{\eta '} - m^2_\eta}
     {m^2_{\eta '} + m^2_\eta -2\mu^2_{88}z^2}\right)^2 - 1}~,
     \label{mixang}
\end{equation} 
where $\mu^2_{88}z^2$ is determined by the pion mass, kaon 
mass and rescaling parameters, i.e.,
\begin{equation}  
    \mu^2_{88}z^2 = \frac{m^2_\pi}{3}
    \left(4\frac{m^2_K}{m^2_\pi}\frac{1}{{z}^2_K}-1\right) {z}^2~.
    \label{aux1}
 \end{equation}
The rescaling parameter $z$ must not be considerably smaller than $z_K$.
Indeed, from Eqn. \ref{massk} for the kaon mass and Eqns. \ref{etamass}
for the $\eta$ and $\eta '$ masses we may write, 
\begin{equation}
	\frac{4}{3}\frac{m^2_K}{m^2_\eta}\frac{1}{z^2_K} - 
	\frac{1}{3}\frac{m^2_\pi}{m^2_\eta}\geq \frac{1}{z^2}~,
	\label{¥}
\end{equation}
or equivalently, $1.03z\geq z_K$. Thus $z$
is at most a factor of $\sim 3\%$ smaller than $z_K$. With the
$Alternative ~I$ parameter set of Table \ref{fit}, the 
rescaling parameters and mixing angle are, 
\begin{equation}
\quad {z}_K = 0.87\pm0.04, 
\quad {z} = 0.89\pm0.06, \quad \theta_P =-(14.3\pm2.2)^o~.
\label{res1}
\end{equation}

How significant are the departure of these parameters from their values
at the exact limit of $U(3)_L\bigotimes U(3)_R$ symmetry?
Clearly, the exact $SU(3)_F$ symmetry limit values 
$c_W =0,~ c_A = d_A = 0,~r = 1$ would be inconsistent with the
measured value of the ratio 
$\Gamma (K^{*0}K^{0}\gamma)/ \Gamma (K^{*+}K^{+}\gamma)$; 
with $c_W = 0$ this ratio becomes 4 as opposed to the experimental value
of $2.34 \pm 0.43$. In addition, direct symmetry breaking alone is not
sufficient; with $c_A = 0$ and $c_W = - 0.2$, the calculated width 
$\Gamma (K^{*0}K^{0}\gamma)$ is about 30$\%$  higher than 
experimental value, well beyond the measurement accuracy. Based on fit 
quality  both direct (i.e. $c_W \neq 0$) and  indirect (i.e. $c_A,d_A \neq 
0, r\neq1$)  symmetry breaking terms are needed to explain data. 

As a further check we may estimate these scales using data directly.
First, from the ratio,
 \begin{eqnarray}
         &&\frac{\Gamma (K^{*0}K^0\gamma)}{\Gamma (K^{*+}K^+\gamma)} =
         4\left[\frac{1+\frac{1}{2}c_W }{1-c_W}\right]^2 =
         \frac{(117\pm10)KeV}{(50\pm 5)KeV} = 2.34\pm 0.43~,
        \label{cwpar}
 \end{eqnarray}
one obtains $c_W = -0.19 \pm 0.04$ in full agreement with the 
global fit results.
Next, the decay width $\Gamma(K^{*0}K^0\gamma)$ involves the kaon
rescaling parameter (or equivalently the scales $c_A$). One finds,  
${z}_K = 0.86 \pm 0.08$ and $c_A=0.52\pm 0.22$.  
The other two parameters are deduced from the ratios, 
$\Gamma(\phi\eta\gamma) / \Gamma(\omega\eta\gamma)$ 
and, $\Gamma(\phi\eta\gamma)/ \Gamma(\eta' \rho \gamma)$, which yield,
   \begin{equation}
         d_A = -0.45 \pm 0.2~, \quad r=0.98\pm 0.1~.
        \label{set1}
   \end{equation}  
With this set of parameters the corresponding mixing angle $\theta_P
= -(16.2\pm2.4)^o$ and $\chi^2/dof = 36/6$ is more than a factor 
of 10 higher as compared to the $Alternative ~ I$  fit results. A global 
fit to data with the assumption of  an exact nonet symmetry, i.e. $r=1$, gives  
$c_W=-0.18\pm0.05,~c_A=0.64\pm0.06,~d_A= -0.32\pm0.04$
and $\theta_P=-(15.2\pm2.2)^o$ but with  $\chi^2/dof=13.2/7$. 
Again the fit quality is significantly poorer as compared to the 
$Alternative ~I$ results. 

It is worthy of mention that the mixing angle is very sensitive to the  
ratio of the rescaling parameters $z$ and $z_K$. The smallest mixing 
angle is obtained with $z = z_K$ (i.e., $d_A = -c_A/4$) and 
readily grows for increasing $z/z_K$. However, a global fit 
with these parameters taken to be equal is again far poorer with  
$\chi^2/dof=8.2/7$ and $c_W=-0.26\pm0.05,~c_A=0.56\pm0.06,
~r=0.86\pm0.1,~\theta_P= -(8.2\pm2.0)^o$. The mixing angle is 
far less sensitive to the direct symmetry breaking scale $c_W$.
Based on these global fit analyses we may conclude that nonet symmetry 
breaking and field rescaling parameters depart slightly but $significantly$ 
from unity.     

Finally, by comparing Eqn. \ref{aux1},  with the familiar quark model
result \cite{pdg00},
\begin{equation}
        m^2_{88} = \frac{m^2_\pi}{3}
        \left(4\frac{m^2_K}{m^2_\pi}-1)\right)(1+\Delta)~,
        \label{}
\end{equation}
the symmetry breaking measure is, 
\begin{equation}
        \Delta \approx \frac { {z}^2}{{z}^2_K} -1~.
        \label{}
\end{equation}
The $Alternative~ I$ parameters correspond to $\Delta = 0.05\pm0.01$.

\underline{{\bf The $Alternative ~II$~~-Two mixing angle scheme}}~~
The mixing angles and rescaling parameters corresponding to the 
$Alternative ~II$ fit parameters of Table \ref{fit} are : \\ 
$\lambda =(25.1\pm2)^o,~\chi=-(34.9\pm3)^o, 
~z_K=0.95\pm0.06,~z=1.00\pm0.06,~f=0.88\pm0.06$~,
or equivalently,
\begin{equation}
\theta_{\eta}=-(5.8\pm2)^o,~~\theta_{\eta'}=-(12.7\pm2)^o,
~~z_1= 0.98\pm0.05,~~z_2=0.96\pm0.05~.  
        \label{}
\end{equation}
In view of the very poor fit quality (confidence level of less than 0.05), 
it seems quite unjustified to use a two mixing angle scheme in
analyzing radiative decays. 

\underline{{\bf Quark flavor basis scheme}}~~
As indicated already, this scheme corresponds to the 
$Alternative ~II$ in the limit of exact nonet symmetry ($r=1$). 
In this case the values of $c_A$ and $d_A$ are considerably 
different, and the rescaling parameters and mixing angle are,
\begin{equation}
\phi = (40.0 \pm 2.0)^o, ~~{z}_K = 0.76\pm0.04,~~z_s=0.88\pm0.05~.
\end{equation}
With the standard octet-singlet mixing angle defined as $\theta_P = \phi
-\theta_{ideal} $ as in Ref.\cite{feld98} we have $\theta_P = -(14.7 \pm
2.0)^o$. This value agrees remarkably well with the $Alternative ~I$ results.  
Yet, with $\chi^2/dof = 19.8/7$ (confidence level less than 0.01) this 
scheme like the $Alternative ~~II$ is far poorer than the $Alternative ~I$.

It is of interest to note that similar mixing angle values were
extracted by several authors using various phenomenological models. Here
we mention few examples.                                       
Feldmann et al.\cite{feld98,feld98a} using their QFB analysis reported a
value $\theta_P = -(14.8 \pm 2.9)^o$. Cao and Signal \cite{cao99} obtained
value of $\theta_P=-(14.5\pm2.0)^o$ from analyzing  large             
momentum  $e^+e^- \to \eta,\eta'\rightarrow 2 \gamma$.
Somewhat a less negative value,
$\theta_P=-(11.59\pm0.76)^o$, was determined
by Benayoun et al. \cite{benayoun99,benayoun991}. The fact that different
model analyses predict similar results should not be
surprising since the mixing angle is sensitive to ratios of scaling
parameters (see Eqn. \ref{mixang}) rather than to their actual 
values. In fact, since this ratio is close to one, the mixing angle
should be determined rather accurately from the experimental 
meson masses. This also means that in similar analyses where 
rescaling is neglected or the corresponding ratios of parameters are 
close to unity, as in our $Alternative ~I$ case, similar values of $\theta_P$ 
are expected. However as our analyses proves, rescaling plays an important 
role in explaining the decay widths.

\bigskip

\section{Summary and Discussion}¥

In this paper, we have constructed a local $U(3)_L \bigotimes U(3)_R$
symmetry based effective field theory, using a generalized
version of the HLS approach of Bando et al.\cite{bando} and a general procedure 
of including the $\eta '$ into a chiral theory\cite{gasser85,leut96,leut97,herera97}. 
At lowest order the lagrangian comprises two terms $L_A$, $L_V$ 
describing respectively, the interactions of pseudoscalar and vector meson 
nonets, and a "kinetic" term for the vector mesons. In variance with the BKY 
model\cite{bando} and variants of this model\cite{bramon95,benayoun981}, 
these terms and likewise their asymmetric companions ${\bar L}_A$ and 
${\bar L}_V$ involve terms which are proportional to the square of traces, 
in addition to terms proportional to the trace of vector square and 
axial-vector square. Such terms must be included in any 
$U(3)_L\bigotimes U(3)_R$ symmetry based theory, and play an important 
role in reproducing the actual meson masses.  In a way similar to that 
proposed by Bramon et al. \cite{bramon95} and 
Benayoun et al.\cite{benayoun98,benayoun99}, we have constructed 
"direct" asymmetric companions ${\bar L}_A$ and ${\bar L}_V$ using 
a matrix $B$ which is proportional to quark mass matrix. With this 
choice of $B$, our theory predicts the same ratios of isospin to $SU(3)$ 
symmetry breaking scales as in QCD. Our lagrangian includes also the 
exact form of  symmetry violating mass term as derived by Herera-Sikl\'{o}dy et
al.\cite{herera97}. Two alternative forms of symmetry breaking parts were 
proposed, the first of $Alternative ~I$ breaks $SU(3)_F$ only while the 
second $Alternative~II$ breaks $U(3)_F$. Each of these alternatives 
corresponds to a unique state mixing scheme, which can be
derived through the diagonalization procedure of the corresponding kinetic
and mass terms, a procedure which turns to be very useful in constraining
the theory parameters.  A third scheme corresponding to the QFB of
Feldmann et al.\cite{feld98},  is realized from the $Alternative~~II$ by 
requiring nonet symmetry. As a result of the diagonalization of 
the kinetic and mass terms of the lagrangian, the mixing schemes, i.e. 
the mixing angles and field rescaling parameters, depend on three symmetry 
breaking scales ($c_A, d_A, r$) only, and by all means can not be 
treated as free parameters. This relation between the mixing schemes and 
the lagrangian form has been overlooked in previous 
studies\cite{benayoun981,benayoun98,gedalin00}.

Using these three schemes we have applied our theory to the pseudoscalar
sector. The case of the vector meson sector is considered
elsewhere\cite{gedalin012}. The results obtained reflect upon the
nature of the symmetry breaking required to explain radiative decay
widths. Based on  global fit analyses, we observe that the $Alternative ~I$ 
provides by far a better description of the data considered. This argues 
for $SU(3)_F$ symmetry breaking supplemented with broken nonet 
symmetry. The other alternatives of $U(3)_F$ symmetry breaking with or 
without nonet symmetry breaking yield rather poor fits and seem unjustified. 

There have been numerous data analyses in the last three decades
attempting to deduce a reliable and accurate value of the pseudoscalar 
mixing angle
\cite{gilman87,bramon90,ball96,bramon97,veno98,benayoun98,bramon99,cao99}.  
The values reported range from 
$\theta = -23^o$ to as high as $\theta = -10^o$. In marked difference with
these previous studies, in the present work the particle mixing scheme
is well related to the lagrangians presumed. All observables are
calculated from the lagrangian with no additional assumptions.
Albeit, we believe that the theory proposed provides an accurate
framework for the study of
electroweak and strong interactions amongst light flavor mesons.

\bigskip

{\bf Acknowledgment}  This work was supported in part by the Israel
Ministry of Absorption.

\begin{table}
\begin{tabular}{cccc}
	\hline
	 &$Alternative~I$ & $Alternative~II$& QFB  \\
	\hline
    $X_\eta$& ${z} \cos\theta_P - \sqrt{2}{r}\sin\theta_P$ 
    & $z\cos\lambda\cos\chi-
    f\sin\lambda\sin\chi +$ &$\sqrt{3}\cos\phi ~,$\\
	
	 && $\sqrt{2}r(-z\sin\lambda\cos\chi-
	 f\cos\lambda\sin\chi)~,$ &    \\
	
     $Y_\eta$ & $-2{z}\cos\theta_P- \sqrt{2}{r}\sin\theta_P$
     & $-2(z\cos\lambda\cos\chi-
      f\sin\lambda\sin\chi) +$ &$-\sqrt{6/\kappa_s}\sin\phi~,$ \\
	
	 && $\sqrt{2}r(-z\sin\lambda\cos\chi-
	f\cos\lambda\sin\chi)~,$ &   \\
	
	$X_\eta'$&  ${z}\sin\theta_P+\sqrt{2}{r}\cos\theta_P$&
	$z\cos\lambda\sin\chi+f\sin\lambda\cos\chi +$
     &  $\sqrt{3}\sin\phi$  \\

	 &&$\sqrt{2}r(-z\sin\lambda\sin\chi+f\cos\lambda\cos\chi)~,$&  \\

	$Y_\eta'$ & $-2{z}\sin\theta_P +\sqrt{2}{r}\cos\theta_P$&
	$-2(z\cos\lambda\sin\chi+f\sin\lambda\cos\chi) +$
    &$\sqrt{6/\kappa_s}\cos\phi~.$ \\
	
	& &$\sqrt{2}r(-z\sin\lambda\sin\chi+f\cos\lambda\cos\chi)~.$&  \\
\end{tabular}
 \caption{ The $X_i$ and $Y_i$ coefficients, Eqn. 
\protect\ref{pnonet}. 
Mixing angles are $\theta_P$ for the $Alternative ~I$ and $\phi$ 
for the QFB scheme. $\kappa_s = 1 + c_A + d_A$ is a matrix element of the
kinetic matrix in the QFB. }
 \protect\label{xyfac}
\end{table}

\begin{table}
 \begin{tabular}{ccc}
	
$v(\rho\pi)=\frac{1}{3}$~ &
$v(\rho\eta)=\frac{1}{\sqrt{3}}X_{\eta}$~ &
$v(\rho\eta') = \frac{1}{\sqrt{3}}X_{\eta'}$~  \\
		
$v(\omega\pi) = 1$~ &
$v(\omega\eta) = - \frac{1}{3\sqrt{3}}X_{\eta}$~ &
$v(\omega\eta') = \frac{1}{3\sqrt{3}}X_{\eta'}$~  \\
		
$v(\phi\pi)=\sqrt{2}¥\epsilon$, & 
$v(\phi\eta) = - \frac{\sqrt{2}¥}{3\sqrt{3}}Y_{\eta}(1+c_W)$ &
$v(\phi\eta') = - \frac{\sqrt{2}¥}{3\sqrt{3}}Y_{\eta'}(1+c_W)$  \\
	
$v(K^{*0}K^0)=v(\bar{K}^{*0}\bar{K}^0)=- 
 \frac{2}{3}{z}_K(1+\frac{1}{2}¥c_W)$ && \\
	
$v(K^{*+}K^+)=v(K^{*-}K^-)=\frac{1}{3}{z}_K(1-c_W)$ & & \\
		
$\bar{v}(\pi)=\frac{3}{\sqrt{2}}$~&
$\bar{v}(\eta)=\frac{1}{\sqrt{6}}\left[ 5X_{\eta}+
         (1+2c_W)Y_{\eta}\right]$~ &
$\bar{v}(\eta')=\frac{1}{\sqrt{6}}\left[ 5X_{\eta '}+
        (1+2c_W)Y_{\eta '}\right]$~ \\
       	\end{tabular}¥
	\caption{The internal symmetry factors $v(VP)$ 
  and $\bar{v(P)}$, 
Eqns. \protect\ref{vpgamma},~~\protect\ref{ptwog}.}
	\protect\label{vfac}
\end{table}¥

\begin{table}
 \begin{tabular}{cccccc}
 \hline
 & $c_W$ & $c_A$ & $d_A$ & $r$ &  $\chi^2/dof$  \\
 \hline
 $Alternative~ I$ &$-(0.20\pm0.05)$ &$(0.64\pm0.06)$
 &$-0.25\pm0.04$ & $0.91\pm0.04$ & $3.1/6$  \\
  $Alternative ~II$&$-(0.27\pm0.05)$&
$0.2\pm0.05$&$0.1\pm0.02$&$0.94\pm0.05$
 & $18.6/6$\\
  QFB &$-(0.19\pm0.05)$ &$(1.4\pm0.1)$ &$-1.1\pm0.1$ &
  *$1$ &  $19.8/7$  \\
 \end{tabular}¥
 \caption{Symmetry breaking scales and $\chi^2/dof$ from global fit
to data. Values marked with an asterisk were kept fixed.  } 
\protect\label{fit}
\end{table}¥

\begin{table}
 \begin{tabular}{ccccc}
 $Decay$ &  $\Gamma_{exp}$(KeV) &  & 
 $\Gamma_{calc}$(KeV)& \\
 &    & $Alternative~I$ & $Alternative~II$ & QFB\\
 \hline
 $\rho\rightarrow \pi\gamma$ & $76\pm10$	& $76\pm10$ &
 $76\pm10$ & $76\pm10$\\
 $\omega\rightarrow \pi\gamma$ & $716\pm43$ 	& *~$716$ &
   *~$716$ &  *~$716$ \\
 $\rho\rightarrow \eta\gamma$ & $36\pm12$ & $40.0\pm4.1$ & 
 $47.4\pm4.2$& $52.8\pm4.5$ \\
 $\omega\rightarrow \eta\gamma$ & $5.5\pm 0.85$ & $5.2\pm0.45$ & 
 $6.1\pm0.5$ & $6.8\pm0.6$ \\
 $\phi\rightarrow \eta \gamma$ & $57.8\pm 1.6$ & $60.8\pm2.4$ &
 $65.7\pm3.5$ & $60\pm3$ \\
 $\phi\rightarrow \eta'\gamma$ & $ 0.30 
  \begin{array}{c}
 			+0.20  \\
 			-0.16
 \end{array}$ & $0.37\pm0.03$ & $0.24\pm0.03$ & $0.4\pm0.05$ \\
 $\eta'\rightarrow \rho\gamma$ & $61.2 \pm 7.5$ & $70.7\pm5.9$ &
 $70.3\pm4.1$ &$78.5\pm6.6$\\
 $\eta'\rightarrow \omega\gamma$ & $6.12 \pm 0.75$ & $6.47\pm0.45$ &
 $6.4\pm0.4$ & $7.2\pm0.5$ \\
 $\pi^0\rightarrow \gamma\gamma$ & $(7.8 \pm 0.55)10^{-3}$
 &*~$7.8\cdot10^{-3}$ & *~$7.8\cdot10^{-3}$ & *~$7.8\cdot10^{-3}$\\
 $\eta\rightarrow \gamma\gamma$ & $0.460 \pm 0.050$  & $0.468\pm0.03$ & 
 $0.58\pm0.02$ & $0.637\pm0.03$ \\
 $\eta'\rightarrow \gamma\gamma$ & $4.280 \pm 0.280$ & $4.02\pm0.3$ & 
 $3.68\pm0.24$ &$4.52\pm0.3$  \\
 $K^{*0}\rightarrow K^0\gamma$ & $117\pm 10$ & $106\pm8.3$ &
 $120.7\pm4.6$ & $113\pm8$  \\
 $K^{*\pm}\rightarrow K^\pm\gamma$ & $ 50 \pm 5$ & $48.6\pm3.2$ &
 $63.3\pm5.2$ &$48\pm5$  \\
 $\chi^2/dof$                      &             & $3.1/6$      &
 $18.8/6$ 	&$19.8/7$ \\
 \end{tabular}¥
 \caption{Calculated radiative decay widths. 
 The $Alternative~I$, $Alternative~II$ and QFB correspond to the parameter
 sets of Table \protect\ref{fit}.
 The widths marked with an asterisk were used to evaluate the coupling
 constants $g_P$ and $g_V$ (see text).
 The experimental decay widths of the second column are
the best fit values of Ref. \protect\cite{pdg00}.}
 \protect\label{vpgpred1}
\end{table}¥

\end{document}